\documentstyle[aps,pre,epsfig,citesort]{revtex}
\twocolumn
\begin{document}

\title{Energy non-equipartition in systems of inelastic, rough spheres}

\author{Sean McNamara$^{(1,2)}$ and Stefan Luding$^{(1)}$}
\address{(1) Institute for Computer Applications 1, University of Stuttgart,
             Pfaffenwaldring 27, 70569 Stuttgart, Germany.\\
         (2) Levich Institute, Steinman Hall T-1M, 140th St and Convent Ave,
             New York, NY 10031, USA}
\maketitle

\begin{abstract}
We calculate and verify with simulations the ratio between the
average translational and rotational energies of systems with rough, 
inelastic particles, either forced or freely cooling. 
The ratio shows non-equipartition of energy. In stationary flows, 
this ratio depends mainly on the particle roughness, but in 
nonstationary flows, such as freely cooling granular media, 
it also depends strongly on the normal dissipation. 
The approach presented here unifies and simplifies 
different results obtained by more elaborate kinetic theories. 
We observe that the boundary induced energy flux plays an important
role.
\end{abstract}

\def\ave#1{\langle #1 \rangle}

\def\D{{\cal D}}
\def\Et{\bar E}
\def\Er{E^\circ}
\def\r{r}
\def\nt{\bar n}
\def\nr{n^\circ}
\def\Pr{P^\circ}
\def\Pcoll{P_{\text{coll}}}
\def\n{\bbox{\hat n}}
\def\bv{\bbox{v}}
\def\t{\bbox{\hat t}}

Granular materials are collections of macroscopic particles
with rough surfaces and dissipative interactions, as addressed in this letter. 
Although rotation and friction are often neglected, they play an 
active role for the dynamics of systems with rough or non-spherical 
constituents. In contrast to classical elastic systems,
energy is not equipartitioned between the degrees of
freedom in the system \cite{jenkins85b,lun87,lun91,Goldshtein95,luding95b}.
In order to examine this ratio, kinetic theories 
\cite{jenkins85b,lun87,lun91,Goldshtein95} and
numerical simulations \cite{luding95b} were applied
for special boundary conditions, and a variety of results were obtained.
We unify these results in a single theory which also explains when each
one is valid.

We consider a system of $N$ particles. We define $\Et$ to be the average
translational kinetic energy per degree of freedom,
and $\Er$ to be the rotational kinetic energy per degree of freedom,
so that:  
\begin{equation}
\Et \equiv {1\over N\nt} \sum_{i=1}^{N} \frac{m}{2} v_i^2, {\rm~and~}
\Er \equiv {1\over N\nr} \sum_{i=1}^{N} \frac{m q}{2} a^2 \omega_i^2.
\label{eq:defE}
\end{equation}
Here, $m$ is the mass and $a$ the radius of a particle. $v_i$ is 
the velocity of particle $i$, and $\omega_i$ is its angular velocity.
$\D$ is the number of dimensions (we restrict ourselves to 
$\D=2$ and $3$ here), $q$ is the dimensionless
moment of inertia; $q=1/2$ for disks, and $q=2/5$ for spheres.
The number of translational degrees of freedom per particle is $\nt=\D$,
and the number of rotational degrees of freedoms is $\nr=2 \D -3$. 
$\Et$ and $\Er$ are often referred to as ``granular
temperatures''.  This terminology is not intended to suggest that a
thermodynamic equilibrium exists in granular flows, but simply to
draw an analogy with the temperature of an ideal
gas, which is also the average energy per degree of freedom.  
$\Et$ and $\Er$ in Eq.~(\ref{eq:defE}) are well defined whether or
not the system is in equilibrium.  In this paper, we consider
all particles to be identical, however, the above definitions can
easily be extended to different types of particles.

We quantify the distribution of energy between the translational and 
rotational modes with the quantity $R \equiv {\Er / (\Er + \Et)}$.
When there is no energy in the rotational mode, $R=0$.  When energy
is equally distributed between all the modes, $R=1/2$.  If rotational
energy dominates, then $R \to 1$.  
We will study how $R$ depends
on the particle properties and the boundary conditions.  This question
has been addressed by several authors
\cite{jenkins85b,lun87,lun91,Goldshtein95}
but serious and unexplained conflicts exist between their results.
For example, \cite{Goldshtein95} claims that $R$ depends on the
normal restitution, whereas \cite{jenkins85b,lun87,lun91}
say $R$ is independent of this parameter.

We use the standard constant roughness model for the 
instantaneous collisions of rotating particles with radius $a$,
mass $m$, and moment of inertia $I=qma^2$.  
This model accounts for dissipation, using the restitution coefficient
$r$ and the tangential restitution $\beta$. Since it has been extensively 
used and discussed \cite{jenkins85b,lun87,lun91,Goldshtein95},
we include only the results here. The post-collisional velocities
$\bv'$, $\bbox{\omega}'$ are given in terms of the pre-collisional
velocities $\bv$, $\bbox{\omega}$ by
\begin{eqnarray}
\bv_{1,2}' &=& \bv_{1,2} \mp {1+\r\over 2} \bv_n \mp
    {q(1+\beta) \over 2q+2} (\bv_t+\bv_r) , {\text{~and}}\cr
a\bbox{\omega}_{1,2}' &=& a\bbox{\omega}_{1,2} + {1+\beta \over 2q+2} 
		\left[\bbox{\hat n}\times(\bv_t + \bv_r)\right],
\label{eq:collrule}
\end{eqnarray}
Here,
$\bv_n \equiv \left [ (\bv_1 - \bv_2) 
    \cdot \bbox{\hat n} \right ] \cdot \bbox{\hat n}$ is the component
of $\bv_1-\bv_2$ parallel to $\bbox{\hat n}$, a unit vector pointing
along the line connecting the centers of the colliding particles.
The tangential component of $\bv_1-\bv_2$ is 
$\bv_t \equiv \bv_1 - \bv_2 - \bv_n$ and
$\bv_r \equiv - a(\bbox{\omega}_1 + \bbox{\omega}_2) \times \bbox{\hat n}$
is the tangential velocity due to particle rotation. 

Later on, we will need expressions for the change in rotational and
translational kinetic energy during a collision.  
The change in translational energy is
\begin{eqnarray}
N\nt\Delta\Et \equiv - Q v_n^2 + S
  \left[-C_{t1} v_t^2 - C_{t2} (\bv_t\cdot\bv_r) + C_{t3} v_r^2\right],
\label{eq:deltaEt}
\end{eqnarray}
with the positive prefactors $Q \equiv m(1-\r^2)/4$, 
$S \equiv mq(1+\beta)/[4(1+q)^2]$, 
and the constants $C_{t1} \equiv 2+q(1-\beta)$,
$C_{t2} \equiv 2-2q\beta$ and $C_{t3} \equiv q(1+\beta)$.
Likewise, the change in rotational energy is
\begin{eqnarray}
N\nr\Delta\Er \equiv + S
  \left[C_{r1} v_t^2 - C_{r2} (\bv_t\cdot\bv_r) - C_{r3} v_r^2\right],
\label{eq:deltaEr}
\end{eqnarray}
where the constants are $C_{r1} \equiv (1+\beta)$,
$C_{r2} \equiv 2(q-\beta)$, and $C_{r3} \equiv 2q+1-\beta$.  
Note that the $C$ are positive (except that $C_{r2}$ can
be negative) so that the signs in Eqs.\ (\ref{eq:deltaEt}) 
and (\ref{eq:deltaEr}) indicate the direction
of energy transfer between the degrees of freedom.

The quantity $R$ (or its equivalent) has been calculated by several
authors.  One result found by three different authors
\cite{jenkins85b,lun87,lun91} for granular material undergoing
uniform shear is  
\begin{equation}
R = {q(1+\beta) \over q(1+\beta) + (1+2q-\beta)},
\label{eq:lun}
\end{equation}
i.e. a function independent of $\r$. 

On the other hand, Goldshtein and Shapiro~\cite{Goldshtein95} found a much
different expression (for $\D=3$), involving $\r$:
\begin{equation}
R = {1\over2}\left(1-{a_G\over b_G + \sqrt{a_G^2+b_G^2}}\right),
\label{eq:Goldshtein}
\end{equation}
with the quantities
$a_G \equiv (1-\beta^2)(1-q)/(1+q) -1+\r^2$, and
$b_G \equiv 2q (1+\beta)^2/(1+q)^2$.
Eq.\ (\ref{eq:Goldshtein}) differs greatly from Eq.\ (\ref{eq:lun})
when $\r<1$.  


In the following, we show that the differences between
Eqs.\ (\ref{eq:lun}) and (\ref{eq:Goldshtein}) arise from the boundary 
conditions, i.e. from the existence of external forcing.



Consider a granular material with external forcing,
where the particles interact only through
collisions obeying Eq.\ (\ref{eq:collrule}). 
The change in $\Er$ from time $t_0$ to $t_1$ is
\begin{equation}
\Er(t_1) - \Er(t_0) = \sum_{\text{coll.}} \Delta \Er({\cal C}_i)
	+ {1\over N \nr} \int_{t_0}^{t_1} \Pr(t)\, {\rm d}t ,
\label{eq:Energybalance}
\end{equation}
where $\Pr$ is the rotational energy added by the forcing.
The sum is taken over all the collisions which take place between
$t_0$ and $t_1$, and $\Delta \Er({\cal C}_i)$ is the change in $\Er$
for collision $i$.
Now, consider a situation where the granular medium is maintained in a
stationary state [$\Er(t_1)=\Er(t_0)$] by some kind of forcing, and
that this forcing adds only translational kinetic energy [$\Pr=0$].  
Eq.\ (\ref{eq:Energybalance}) becomes:
$\sum_{\text{coll.}} \Delta \Er({\cal C}_i) = 0$, which states 
that collisions do not, on average, change the rotational energy.
When the assumptions made above are
satisfied, the dissipation of rotational energy is exactly balanced
by the conversion of $\Et$ into $\Er$.
Using Eq.\ (\ref{eq:deltaEr}) we get:
$- C_{r1} \langle v_t^2 \rangle +
 C_{r2} \langle \bv_t\cdot\bv_r \rangle +
 C_{r3} \langle v_r^2 \rangle = 0$.
Here, the angle brackets $\langle ... \rangle$ indicate an 
average taken over the collisions.  We now consider how to calculate
these averages.

If $\psi$ is the quantity to be averaged, then
\begin{equation}
\langle\psi\rangle = \{ \psi \Pcoll ( \bv_1, \bv_2,
	\bbox{\omega}_1, \bbox{\omega}_2, b) \}_{b,1,2},
\end{equation}
where $\Pcoll$ gives the probability of a collision
occuring between particles 1 and 2 with a normalized
impact parameter $0\le b \le 1$.  The normalized impact parameter 
is the distance between particle centers at closest approach
if the particles did not interact, normalized by the particle diameter.
The subscripts on the brackets means we average over all values of $b$,
and over all pairs of particles.  We now make several simplifying
assumptions about $\Pcoll$.  First, we assume that the angular
velocities have no effect on the probability of collision:
$\Pcoll = \Pcoll(\bv_1,\bv_2,b)$.  Next, we assume that 
$\Pcoll \sim v f(b)$, ($v\equiv \left|\bv_1-\bv_2\right|$) because particles
with large relative velocities are more likely to collide.  Finally,
the dependence of $\Pcoll$ on $b$ can be deduced from geometrical
arguments: $\Pcoll \sim v$ for $\D=2$ and $\Pcoll \sim vb$ for
$\D=3$.  Thus we have
\begin{equation}
\langle\psi\rangle = \{ \psi v (2b)^{\D-2} \}_{b,1,2} /
		\{ v \}_{b,1,2},
\end{equation}
with the factor of $2$ and the denominator required for normalization.

Since the rotational and translational velocities are uncorrelated,
\begin{equation}
\langle v_r^2 \rangle = \{ v_r^2 \}_{1,2} = 2a^2 \left\{(\bbox{\omega}_1 \times
	\bbox{\hat n})^2 \right\}_1 = \frac{4(\D-1)}{qm} \Er.
\end{equation}
The factor of $\D-1$ arises because in $\D=3$, one of the three rotational
degrees of freedom is excluded by the cross product with $\bbox{\hat n}$.
In $\D=2$, there is only one rotational degree of freedom, and it always
participates in every collision.

To calculate $\langle v_n^2 \rangle$ and $\langle v_t^2 \rangle$,
we use $v_t^2 = v^2 b^2$ and $v_n^2 = v^2-v_t^2$.  The average
can be factored into two parts:
\begin{equation}
\langle v_t^2 \rangle = \frac{\{ v^3 b^2 (2b)^{\D-2} \}_{b,1,2}}
		{\{ v \}_{1,2} } = \{ b^2 (2b)^{\D-2} \}_b
	  \frac{\{v^3\}_{1,2}}{\{v\}_{1,2}}.
\end{equation}
Evaluating the first factor gives $\{ b^2 (2b)^{\D-2} \}_b=\D/6$.
The second factor will be proportional to $\Et/m$, but calculating the
coefficient requires knowledge of the distribution of velocities.  Assuming
a Maxwellian velocity distribution gives
$\{v^3\}_{1,2}/\{v\}_{1,2} = 24 (\D-1) \Et / (\D m)$.  Then,
we have $\langle v_t^2 \rangle = 4(\D-1)\Et/m$ and 
$\langle v_n^2 \rangle = 8\Et/m$.  

All of the assumptions we have made up to know are equivalent to those
made in the kinetic theories \cite{jenkins85b,lun87,lun91,Goldshtein95}.  
Thus, it is not surprising that we recover some of their results.  
In particular, putting the averages into Eq.\ (\ref{eq:Energybalance}) 
gives $- C_{r1} \Et + C_{r3} \Er / q = 0$, and after
using the definitions of $C_{r1}$ and $C_{r3}$, we obtain
$R$ as in Eq.\ (\ref{eq:lun}).  

Eq.~(\ref{eq:lun}) does not depend on $r$ because $R$ is
fixed by a balance between the conversion of $\Et$ into $\Er$
and the dissipation of $\Er$. Both of these processes depend only
on $\beta$ and $q$ but not on $r$. As soon as the dissipation of $\Et$
starts to play a role  in determining $R$, then $r$ will appear.
We examine such a case next.\\

Consider a granular medium in the absence of forcing.  If $r\neq1$ and
$\beta\neq1,-1$, then $\Er$ and $\Et$ decrease towards $0$,
but $R$ can approach a constant. This can be verified by simulations
and a more elaborate calculation in the framework of the kinetic theory
\cite{Goldshtein95} or with a Liouville operator formalism
\cite{huthmann97}.

In the following, we simplify the algebra by using
$K = \Et / \Er$ instead of $R$ [$R=1/(1+K)$].
During a collision, $K$ changes by
\begin{equation}
 \Delta K = \frac{\Et + \Delta \Et }{ \Er+\Delta \Er } - K
= \frac{\Delta\Et-K\Delta\Er }{ \Er+\Delta\Er} .
\end{equation}
We look for a value of $K$ such that $\Delta K=0$. 
The denominator of this equation is always positive.  Equating
the numerator to $0$, we find that a collision leaves $R$ unchanged if
$K = \Delta \Et / \Delta \Er $.
This equation can be expanded in terms of $\ave{v_n^2}$, $\ave{v_t^2}$,
and $\ave{v_r^2}$ using Eqs.\ (\ref{eq:deltaEt}) and (\ref{eq:deltaEr}), to
\begin{equation}
K = { A \ave{v_n^2} + C_{t1} \ave{v_t^2} - C_{t3} \ave{v_r^2} \over
	- \alpha C_{r1} \ave{v_t^2} + \alpha C_{r3} \ave{v_r^2}},
\label{eq:Kcool}
\end{equation}
where $\alpha\equiv\nt/\nr$, and $A \equiv Q/S$ [see Eq.\ (\ref{eq:deltaEt})].
Because the energy decreases with every collision, the averages must 
be interpreted as taken over all possible collisions at a given time.  

Using our previous expressions for the averages and reorganizing
Eq.~(\ref{eq:Kcool}) as a quadratic equation for $K$ yields
\begin{equation}
\alpha{b_G \over 2} K^2 - \left[ \alpha a_G+ (\alpha-1)c_G \right] K
	- {b_G\over 2} = 0,
\label{eq:K}
\end{equation}
with the quantities $a_G$ and $b_G$ from Eq.\ (\ref{eq:Goldshtein}),
and $c_G \equiv (1+\beta)(2q+q^2-\beta q^2)/(1+q)^2$.
In deriving Eq.~(\ref{eq:K}), we used $\alpha = 2/(\D-1)$.
For $\D=3$, we have $\alpha=1$, and the solution of Eq.\ (\ref{eq:K}) 
leads to Eq.\ (\ref{eq:Goldshtein}).\\

Next we compare the theoretical results, derived above,
with simulations in $\D=2$.  We examine three different simulational
``experiments''.  In the first case, energy is put into one translational
mode by a vibrating wall, and in the second case, a granular material 
under sheared periodic boundary conditions is examined.
Finally, a granular media is studied in the absence of any forcing
whatsoever.  On the basis of the theory presented above,
we expect the first two cases to obey Eq.\ (\ref{eq:lun}), and the last
case to obey Eq.\ (\ref{eq:Goldshtein}).  
In all cases, we perform the experiments with different $r$, and
for each value of $r$, we vary $\beta$ from $-0.95$ to $1$.

\begin{figure}[htb]
\epsfig{file=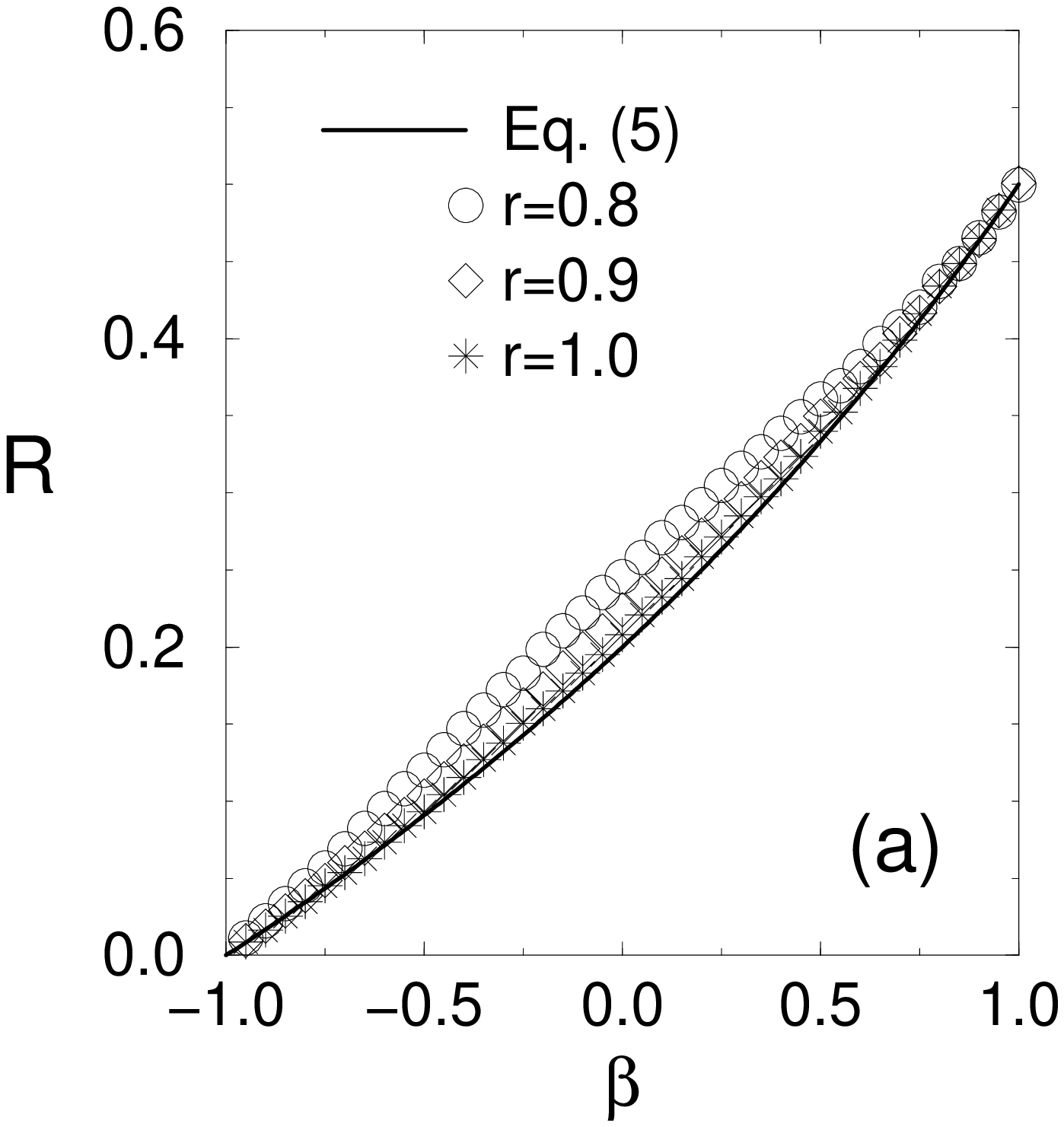,height=4.7cm} ~\hspace{-.0cm}~
\epsfig{file=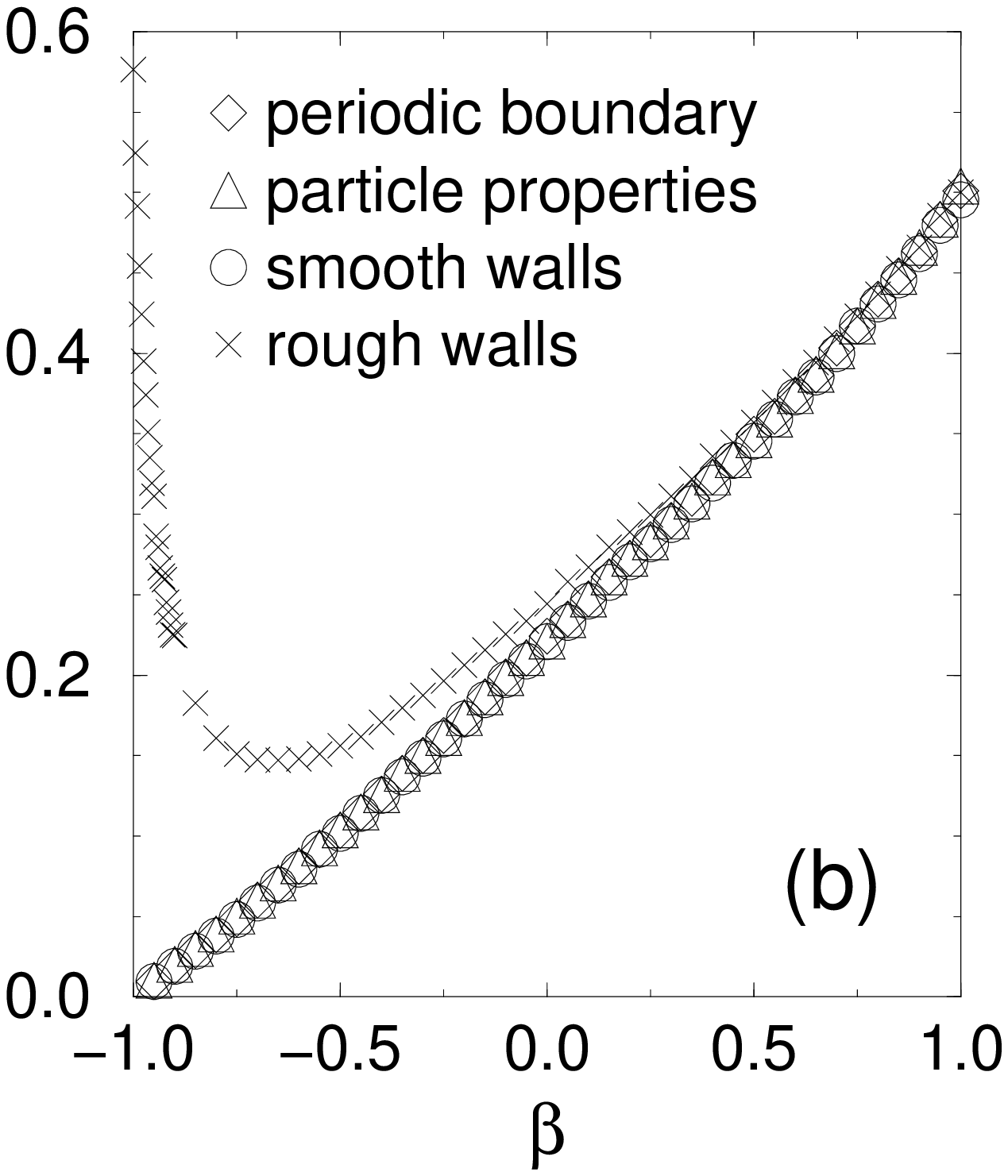,height=5.0cm}
\caption{Simulation results from a vertically vibrated system.
(a) $R$ is compared with the theoretical prediction in Eq.\ 
(\protect \ref{eq:lun}) for three values of $\r$.
(b) The data for $r=0.9$ from Fig.\ \protect{\ref{fig:vibresults}(a)}
are compared to simulations with different boundary conditions
(see text for details). }
\label{fig:vibresults}
\end{figure}

In the first experiment $N=160$ particles of radius $a$ are placed 
on a vertically vibrating floor in the presence of gravity.  
The boundaries in the horizontal direction are periodic,  
the domain is $50$ particle radii wide and infinitely high.
The period of the floor vibration $T$ and the gravitational acceleration
are here related by $gT^2/a=1$.  The height of the
floor varies periodically in time, following an asymmetric 
sawtooth wave form.  (The choice of wave form is arbitrary; changing the
wave form does not significantly change $R$.)
The floor moves up a distance of
$5a$, with upwards velocity $5a/T$ and then returns
instantly to its lowest position.
In all cases, the simulations ran for $7000T$, with
$R$ being measured every $\Delta t=0.25T$ for $5000T \leq t \leq 7000T$,
and these values were averaged to give the points in
Fig.~\ref{fig:vibresults}(a).
Since this experiment satisfies the assumption that energy is input into
the translational modes only, we expect that the results will
satisfy Eq.\ (\ref{eq:lun}).  
Fig.~\ref{fig:vibresults}(a) confirms that this is indeed the case,
besides some systematic underestimation of $R$ by the theory
when the dissipation is strong. 

Since it is difficult to do experiments with periodic boundaries,
we also examined the effect of stationary side walls at the edges of the domain.
We compare the results for $r=0.9$ from Fig.\ \ref{fig:vibresults}(a)
with three different types of boundary conditions. 
(i) a perfectly rough wall with $\r=\beta=1$, 
(ii) a wall where $\r$ and $\beta$ have the same values for 
     particle-particle and particle-wall collisions, and 
(iii) a perfectly elastic and smooth wall with $r=1$ and $\beta=-1$.  
The results presented in Fig.~\ref{fig:vibresults}(b) show that only 
the first type of wall causes $R$ to deviate significantly.  
We relate this to a competition between
particle-particle and particle-wall collisions:  
The collisions with the wall in case (i) push $R$ towards $R(\beta=1)=1/2$, 
i.e. equipartition holds, and the collision between particles push 
$R$ towards a smaller value.  (At $\beta_p=-1$, $R>1/2$ because the kinetic
energy of the vertical velocities is greater than that of the horizontal,
and the wall couples only to the vertical motions.)
In case (ii), both types of 
collisions push $R$ towards the same value, and in case (iii),
the particle-wall collisions do not influence $R$,
so that the results are not perturbed at all.

In the second experiment, we drive the granular material by shearing it,
a case frequently considered in the literature \cite{jenkins85b,lun87,lun91}.
$N=160$ particles are placed in a square 
domain whose sides are $L=50a$ in length.
The boundaries are periodic in the $x$ and $y$ direction.  A uniform shear
is imposed by applying Lees-Edwards boundary conditions \cite{AllenTildesley}:
when a particle
exits the domain at the bottom (top), its image enters at the top (bottom)
of the domain, with its $x$ velocity increased by a constant velocity $U$
($-U$), at its position shifted by a distance $Ut$ ($-Ut$).  
These boundary conditions eliminate the need to specify wall
properties and make the system translationally invariant.  
In our simulations, $U=2a/T$, giving a shear rate of $\Gamma = U/L = 0.04/T$.
The time unit $T$ is arbitrary.

\begin{figure}[htb]
\epsfig{file=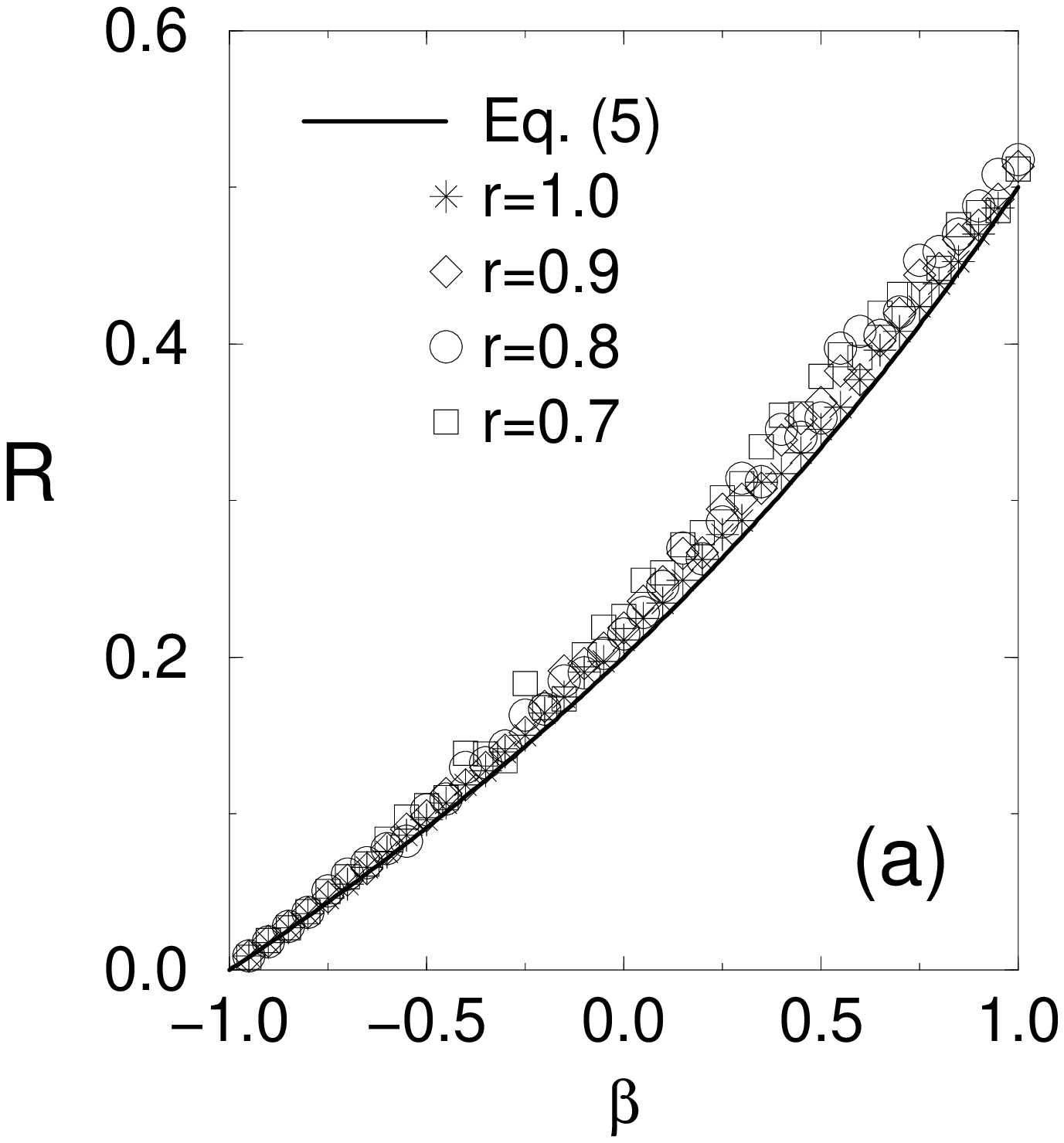,height=4.8cm} ~\hspace{-.0cm}~
\epsfig{file=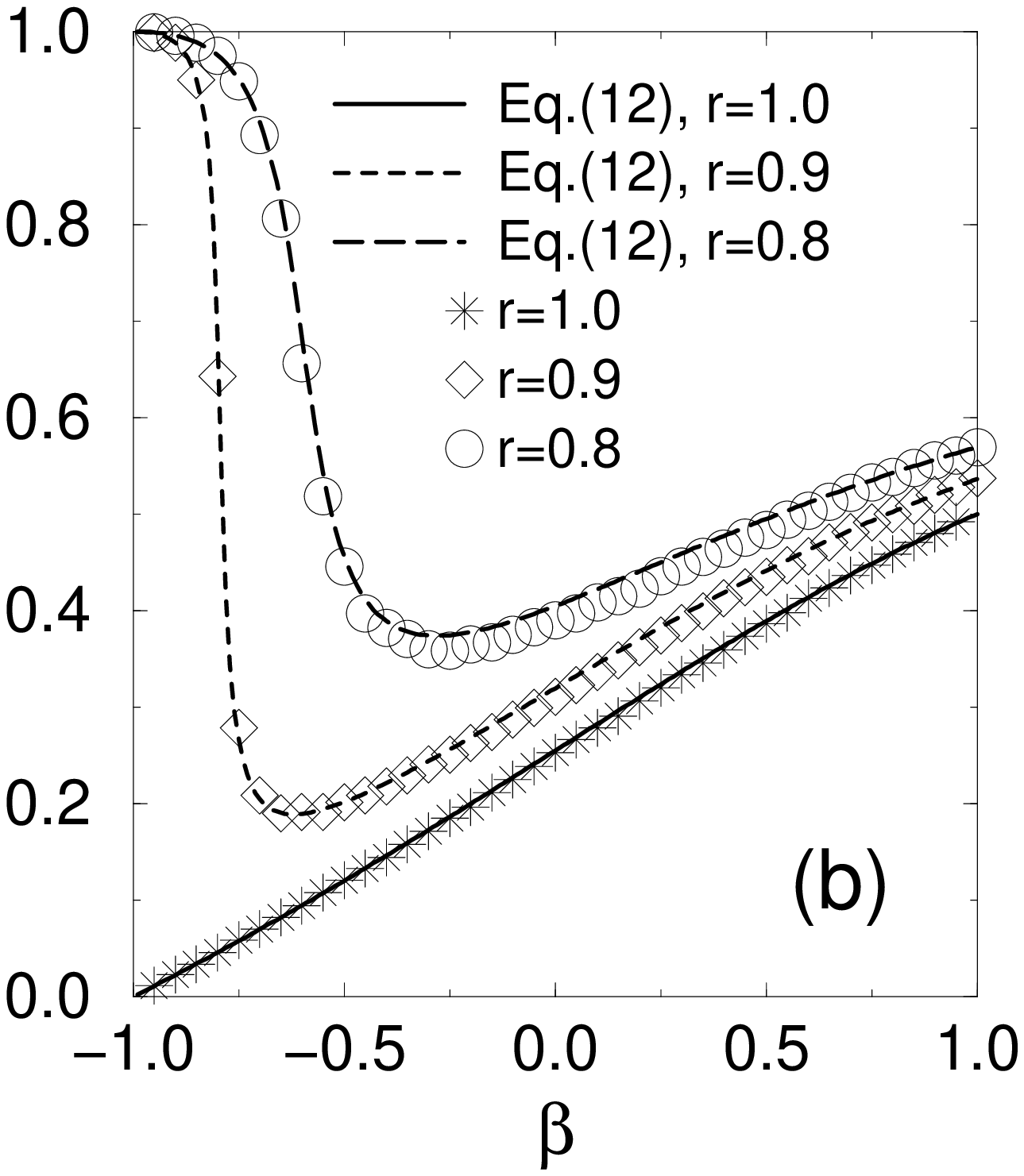,height=4.8cm}
\caption{
(a) Results of the shearing experiment. The points show the results
    for different $\r$, compared to the analytical result.
(b) Results of the cooling experiment. The points show the results of 
simulations and the curves give $R=1/(1+K)$ with $K$ from 
Eq.\ (\ref{eq:K}), with the appropriate value of $\r$, and $\D=2$.}
\label{fig:shearresults}
\label{fig:coolresults}
\end{figure}

A direct application of Eq.\ (\ref{eq:lun}) will fail, because
the boundary conditions generate both an average flow and an average
rotation.  However, if we interpret $\Er$ and $\Et$ to be the energy
which remains after removing the mean flow and rotation, the
agreement between theory and simulation is good, as shown in
Fig.~\ref{fig:shearresults}(a).  The translational kinetic energy
of the mean flow was calculated as
$\Et_{\text{mean}} = (m/2) \Gamma^2 \sum_{i=1}^N (y_i-L/2)^2.$
The rotational energy of the mean flow was estimated by
$\Er_{\text{mean}} = mqa^2 \Omega^2/2$,
where $\Omega$ is the observed average angular velocity.
Even with longer averaging times
($5000T$ in our simulations), the data is considerable noisier than
in Fig.~\ref{fig:vibresults}. 

In the last experiment, $N=160$ particles are placed in a 
periodic domain with $L=100a$, but no shear is applied.  
The initial condition is generated by setting
$r=1$ and $\beta=-1$.  The particles quickly attain a Maxwellian velocity
distribution.  Then, dissipation is ``switched on'', and the system
evolves without any further input of energy.  Although the energy
decreases with every collision, $R$ approaches the constant value
shown in the graph.  The results of many simulations were averaged
together to reduce the fluctuations.
In Fig.~\ref{fig:coolresults}(b), we plot the results.   The theoretical
curves are the solutions of Eq.\ (\ref{eq:K}), with $\alpha=2$, and
the three values of $\r$ shown in the plot.\\

In conclusion, we extend Eq.~(\ref{eq:Goldshtein}) to two
dimensions and summarize the different results for $R$ in the
literature.
Our method of calculation is simple enough to show why Eq.~(\ref{eq:lun})
applies to forced granular media, and Eq.~(\ref{eq:Goldshtein})
to cooling granular media.  Eq.~(\ref{eq:lun}) will apply whenever
the forcing adds only translational energy.  This is the case for
both vibration and shear.
We succeeded to formulate a procedure to
calculate the ratio of tangential and rotational energy in
dissipative systems. Our result replaces the value
$K=1/2$ (in 2D when equipartition is true for elastic systems).
We examined not only the limit of almost elastic particles,
but also rather inelastic situations. For most boundary conditions
used, the agreement between theory and simulations is encouraging.
However, simulations in $\D=3$ are stil needed to check
the analytical expressions.  Possible future work includes
boundary conditions with with non-zero rotational energy input,
and calculating $R$ for the more realistic interaction model that
accounts also for Coulomb-friction. \cite{Foerster94,Walton86,luding95b}

We acknowledge the support of the ``Alexander von Humboldt-Stiftung'' and of
the ``Deutsche Forschungsgemeinschaft, Sonderforschungsbereich 382''.

~\vspace{-0.5cm}~


\end{document}